\documentclass[12pt]{article}

\usepackage{newtxtext,newtxmath}

\usepackage{graphicx}
\usepackage{caption}
\usepackage[letterpaper,margin=1in]{geometry}

\renewenvironment{abstract}
	{\quotation}
	{\endquotation}

\date{}

\makeatletter
\renewcommand{\fnum@figure}{\textbf{Figure \thefigure}}
\renewcommand{\fnum@table}{\textbf{Table \thetable}}
\makeatother

\usepackage{scicite}

\usepackage{url}

\def\scititle{
	\textbf{Stimulated Brillouin Scattering in InGaP-on-Insulator Waveguides}
}
\title{\bfseries \boldmath \scititle}

\author{
	Yuyang~Xue$^{1,2,3,\ast}$,
	Lisa-Sophie~Haerteis$^{2,3,4,5}$,
    Ryan~L.~Russell$^{4,5}$,
    Choon~Kong~Lai$^{4,5}$,\and
    Benjamin~J.~Eggleton$^{4,5}$,
    Michael~J.~Steel$^{6}$,
    Glenn~Solomon$^{3,7}$,
    Kevin~L.~Silverman$^{1}$,\and   
    Moritz~Merklein$^{4,5}$,
    Andreas~Boes$^{2,3}$,
    Nima~Nader$^{1}$\and
	\small$^{1}$Applied Physics Division, National Institute of Standards and Technology, Boulder, CO 80305, USA.\and 
    \small$^{2}$School of Electrical and Mechanical Engineering, Adelaide University, Adelaide, SA 5005, Australia.\and
	\small$^{3}$Institute for Photonics, Advanced Sensing and Quantum Technologies (IPAS-QT), Adelaide University,\and \small
    Adelaide, SA 5005, Australia.\and
    \small$^{4}$Institute of Photonics and Optical Science (IPOS), School of Physics, The University of Sydney,\and \small Sydney, NSW 2006, Australia.\and
    \small$^{5}$The University of Sydney Nano Institute (Sydney Nano), The University of Sydney,\and \small Sydney, NSW 2006, Australia.\and
    \small$^{6}$School of Mathematical and Physical Sciences, Macquarie University, Sydney, NSW 2109, Australia.\and
    \small$^{7}$Department of Physics, Adelaide University, Adelaide, SA 5005, Australia.\and
	\small$^\ast$Corresponding author: yuyang.xue@nist.gov,  yuyang.xue@adelaide.edu.au \and
}

\begin{document} 

\maketitle

\begin{abstract} \bfseries \boldmath

Stimulated Brillouin Scattering (SBS) is a nonlinear interaction between optical and acoustic waves in solids.
First regarded as a parasitic process in optical fibers, it has gathered significant interest in microwave photonic applications such as optical signal processing and narrow linewidth lasers.
While many Brillouin system demonstrations have been done on photonic chips, they struggle to provide simultaneous high Brillouin gain and narrow linewidth in a scalable platform, capable of integration with established photonic integrated circuits. 
Here, we present a novel integrated InGaP-on-SiO\textsubscript{2} platform, where a single crystalline InGaP waveguide layer offers superior material properties and strong nonlinearities to support SBS near 1550~nm wavelength. 
We demonstrate backward SBS with a measured high Brillouin gain coefficient of 588~W\textsuperscript{-1}m\textsuperscript{-1} and a record narrow linewidth of 5.2~MHz at 9.346~GHz frequency shift. 
This work establishes a wafer-scale complementary metal-oxide semiconductor compatible fabrication process, paving a scalable path for high gain and narrow linewidth Brillouin photonics with applications such as precision signal processing.


\end{abstract}


\noindent
\section*{INTRODUCTION}

Photonic integrated circuits (PICs) offer a scalable route to realize photonic systems with superior performance metrics when compared to their free space or optical fiber based counterparts. The use of high refractive index photonic materials enables strong optical mode confinement, allowing for a dramatic footprint reduction and system miniaturization. The resulting system size and weight reduction has rendered PICs as promising chip-scale platforms for photonic systems that utilize linear and nonlinear optical interactions. Integrated photonics also enable higher optical field intensities and enhanced nonlinear optical interactions that can result in increase system efficiency and reduced power consumption \cite{Burla:14,Zhuang:15,Perez2017,Annoni2017,Ribeiro:16,Harris2017,Shekhar2024}.

PIC-based opto-mechanical architectures offer a transformative integration of photonics and mechanical systems at nanoscale, enabling unprecedented control over photon-phonon interactions \cite{RevModPhys.86.1391,VanLaer2015}.
Stimulated Brillouin Scattering (SBS) is one of the most compelling and versatile opto-mechanical interactions realizable on a PIC platform that provides powerful solution for optical filtering, sensing, computing, communication, and data storage \cite{Eggleton2013-gn,Eggleton2019-fa}. SBS arises from a coherent interaction between phonons and photons, where two (counter-) propagating optical waves (pump and probe) resonantly excite and are scattered by acoustic phonons. The scattered light typically has a narrow linewidth on the order of tens of megahertz.
Furthermore, the co-localization of optical and acoustic modes within the micron-scale waveguides produces a high acousto-optic modal overlap, making the on-chip systems highly efficient photonic-phononic transducers. This offers a promising platform for revolutionizing a variety of integrated photonic applications, involving narrow linewidth microwave photonic filters, lasers, acousto-optic data storage, high-resolution sensing and optical gyroscopes \cite{Byrnes:12,8302452,Gundavarapu2019,Choudhary:17, Merklein2017,Zarifi:19,Li:17,Gundavarapu:18}. 


A central challenge in integrated Brillouin systems lies in achieving photonic platforms with low linear and nonlinear losses to enable high Brillouin gain and narrow linewidth operation. Such a platform is also required to be compatible with the existing PIC architectures for scalability and  system integration. Materials with high refractive index, high optical nonlinearity, and those allowing efficient acoustic confinement are critical to enhance SBS interactions. On-chip SBS has been demonstrated in various material platforms such as chalcogenide glasses (ChGs) \cite{Pant11,Morrison:17,https://doi.org/10.1002/adfm.202105230,10.1063/5.0220496,9431687,Lai:23}, silicon nitride \cite{BOTTER2022}, silicon \cite{VanLaer2015,Lei2024} and thin-film lithium niobate (TFLN) \cite{doi:10.1126/sciadv.adv4022,10.1063/5.0274854}. Group III-V semiconductors integrated on oxidized silicon wafers \cite{10.1063/5.0098984,10.1063/5.0225747,Ahler:25,Lin201800149,Stanton:20,Jin:26} are particularly appealing for on-chip SBS due to their high $\chi$\textsuperscript{(2)} and $\chi$\textsuperscript{(3)} nonlinearity, high refractive index, single crystalline structures, large bandgap, and their scalable, complementary metal-oxide semiconductor (CMOS) compatible fabrication process, making them suitable for system-level integration with the established PIC architectures \cite{Dave:15,Akin2024,10.1063/1.5122775}.  

In\textsubscript{0.48}Ga\textsubscript{0.52}P (noted as InGaP from here on), in particular, can be epitaxially grown lattice-matched to GaAs substrates to provide a high optical quality single crystalline semiconductor layer.
InGaP has a wide bandgap of 1.9~eV, suppressing two photon absorption losses near 1550~nm. It also exhibits high $\chi$\textsuperscript{(2)} and $\chi$\textsuperscript{(3)} coefficients of 220~pm/V and $1.1 \times 10^{-13}$~cm$^2$/W, respectively \cite{Baboux:23,10.1063/5.0225747}. Further, compared to Al- and As-based III-V compounds, it does not suffer from optical losses that are due to defect states \cite{Kaminska_1987}, oxidation \cite{Parrain:15}, surface reconstruction layer \cite{Parrain:15,Stanton:20}, and As-As bonds \cite{Stanton:20}.


In this work, we utilize the InGaP-on-SiO\textsubscript{2} platform \cite{10.1063/5.0225747} to demonstrate backward SBS with an ultra-narrow linewidth of 5.2~MHz at 9.346~GHz Brillouin frequency shift, providing a mechanical quality factor of 1797.3 and a calculated acoustic decay time of 61.2~ns. 
Our results show a Brillouin gain coefficient, $g$\textsubscript{B}, of 588~W\textsuperscript{-1}m\textsuperscript{-1} and we achieve a Brillouin gain of 6.6~dB at 156~mW on-chip pump power in a 49.8~mm long waveguide. 
Furthermore, we demonstrate the scalability of this InGaP-on-SiO\textsubscript{2} platform using a 76.2~mm wafer-scale direct bonding process \cite{Nader:25}. 
Based on the ultra-high Brillouin gain coefficient and record narrow linewidth that are competitive against existing platforms, InGaP arises as a strong candidate for Brillouin integrated photonics, \textit{e.g.}, on-chip Brillouin lasers, ultra-narrow filters and photonic-phononic memories \cite{Eggleton2019-fa}.

\clearpage

\section*{RESULTS}
\subsection*{Integrated InGaP-on-SiO\textsubscript{2} platform for Brillouin photonics}

\begin{figure}[ht] 
	\centering
	\includegraphics[width=0.95\textwidth]{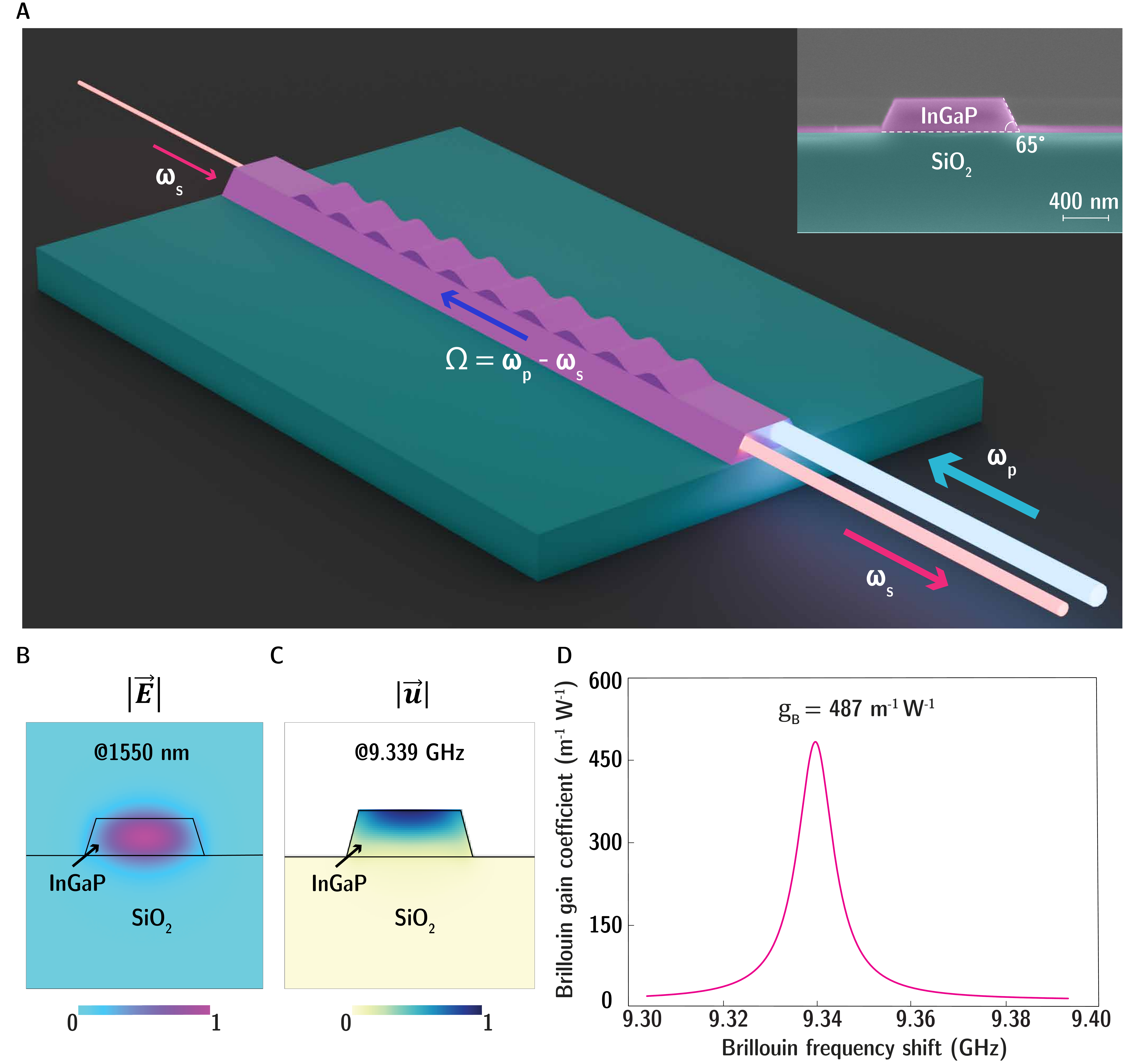}
	\caption{\textbf{Backward SBS in InGaP waveguides.}
	  (\textbf{A}) Schematic illustration of waveguide SBS, with optical probe (red, top left), pump (blue, bottom right) and scattered Stokes wave (red, bottom right). The inset of (A) shows a false-color scanning electron micrograph (SEM) of the waveguide cross section. (\textbf{B}) Quasi-TE\textsubscript{0} mode at 1550~nm. (\textbf{C}) Mechanical displacement profile at 9.339~GHz. (\textbf{D}) Simulated Lorentzian SBS peak for 1000~nm wide and 250~nm tall waveguide, with an assumed mechanical quality factor of 1800, close to the measured value.}
	\label{Figure0} 
\end{figure}


Figure~\ref{Figure0}A illustrates backward SBS in an InGaP-on-SiO\textsubscript{2} straight waveguide with $65^\circ$ sidewall angle, as shown in the inset.
The frequencies of optical pump and probe waves are at $\omega$\textsubscript{p} and $\omega$\textsubscript{s}, respectively. When detuning the probe from the pump by the Brillioun shift, defined as $\Omega = \omega_{\mathrm{p}}-\omega_{\mathrm{s}} \approx 2\frac{n_{\mathrm{eff}}\omega_{\mathrm{p}}}{c}v_\mathrm{a}$, with $n$\textsubscript{eff} the effective index of the optical mode at the pump wavelength, $c$ the speed of light, and $v$\textsubscript{a} the acoustic velocity in the waveguide core \cite{Wolff2021-br}, the counter-propagating pump and probe beams will generate an acoustic wave at the frequency of the Brillouin shift. The generation of this acoustic wave results in a back-scattered Stokes wave, with its frequency downshifted by $\Omega$ compared to the pump wave.

We simulate backward SBS pumped at 1550~nm (Fig.~\ref{Figure0}B) in a 1000~nm wide InGaP-on-SiO\textsubscript{2} waveguide with the fabricated trapezoidal cross-section. InGaP waveguides benefit from a refractive index of 3.11 at 1550~nm \cite{Duan2025} for strong optical mode confinement in the waveguide core. This results in a fundamental TE mode (Fig.~\ref{Figure0}B) with effective refractive index of $n$\textsubscript{eff} = 2.1.
In addition, InGaP has lower longitudinal and transverse bulk acoustic velocities of 5218~m/s and 3608~m/s, when compared to those of the SiO\textsubscript{2} bottom-cladding (5968~m/s and 3764~m/s, respectively) \cite{Goldberg1999GaInP,Haynes2016CRCSection14}, enabling acoustic mode confinement in the waveguide structure. 
We also note that in our simulations, a waveguide structure with air top-cladding provides higher Brillouin gain coefficient when compared to similar structures with SiO\textsubscript{2} top-cladding.
Figure~\ref{Figure0}C shows the mechanical displacement profile for one of the strongest acoustic modes, where the acoustic mode profile is similar to those found in TFLN waveguides~\cite{10.1063/5.0274854}, at the SBS resonance frequency of 9.339~GHz, with an estimated Brillouin gain coefficient of 487~W\textsuperscript{-1}m\textsuperscript{-1} (Fig.~\ref{Figure0}D).

\clearpage

\subsection*{Waveguide fabrication and characterization}

\begin{figure}[ht] 
	\centering
	\includegraphics[width=0.95\textwidth]{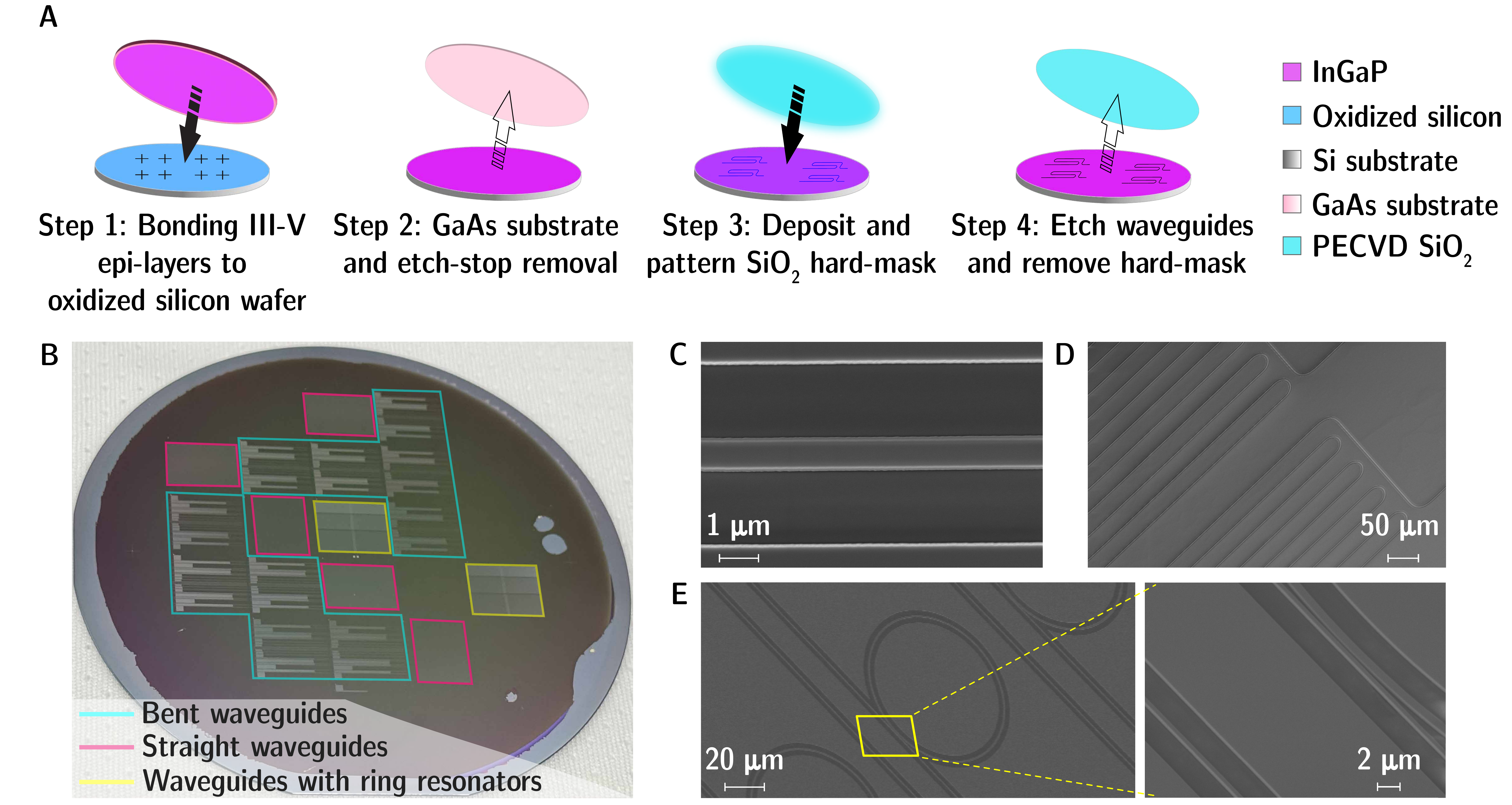}
	\caption{\textbf{InGaP-on-SiO\textsubscript{2} waveguide fabrication.}
	  (\textbf{A}) Illustration of the waveguide fabrication process, including the wafer bonding step. (\textbf{B}) Camera image of 17 chips on a 76.2~mm diameter wafer with 3 different types of designs. (\textbf{C}) SEM of a straight waveguide section. (\textbf{D}) SEM of waveguide bent sections. (\textbf{E}) Bus-waveguide coupled ring resonators (left), and the ring-bus coupling section (right).}
	\label{Figure1} 
\end{figure}

We design and fabricate the InGaP-on-SiO\textsubscript{2} waveguides based on a 250~nm thick single crystalline InGaP layer grown lattice-matched on a 76.2~mm diameter GaAs handle wafer using molecular beam epitaxy.
Figure~\ref{Figure1}A presents an overview of the InGaP-on-SiO\textsubscript{2} fabrication process, involving wafer bonding, electron-beam lithography (EBL) and wet-chemical etching with more details provided in Materials and Methods.
Figure~\ref{Figure1}B shows a camera image of the fabricated 76.2~mm diameter wafer, with three types of chip designs highlighted by different colors. These are straight waveguides for etch geometry calibration (Fig.~\ref{Figure1}C), paperclip structures \cite{Orcutt:12} of various widths and lengths for SBS measurements (Fig.~\ref{Figure1}D), and waveguide-coupled ring resonators for resonator quality factor testing and optical loss characterization (Fig.~\ref{Figure1}E). More details regarding specific chip designs can be found in Materials and Methods.\\

\clearpage

\begin{figure}[ht] 
	\centering
	\includegraphics[width=0.95\textwidth]{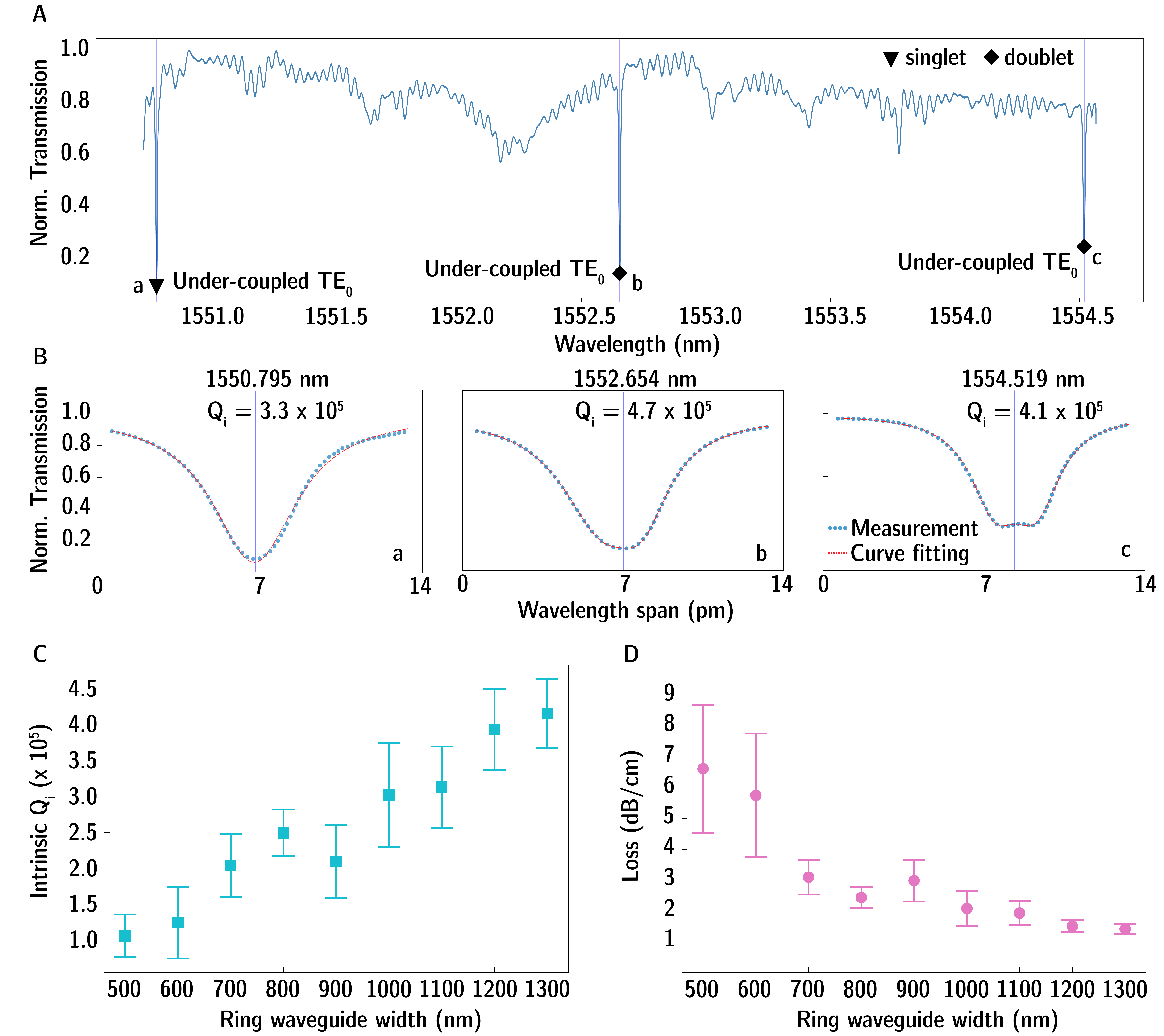}
	\caption{\textbf{Ring resonator measurement and resonance analysis for the fundamental TE mode.}
	  (\textbf{A}) Transmission spectrum of a waveguide-coupled ring resonator. (\textbf{B}) Three resonance peaks a,b and c are plotted with curve fitting. (\textbf{C}) Average intrinsic Q-factors with standard deviation as error bars and (\textbf{D}) correspondingly calculated propagation losses plotted against ring waveguide width.}
	\label{Figure2} 
\end{figure}

To characterize the optical quality factors (noted as Q-factors from here on) of the ring resonators, we couple the output beam of a tunable single-mode laser into a lensed fiber with a focused-spot mode-field diameter of 2~$\mu$m. The focused output of the lensed fiber is coupled to the TE\textsubscript{0} mode of the straight bus-waveguides through 194~nm wide inverse taper structures with estimated coupling efficiency of -3.8~dB per facet. 
An in-line fiber-based unbalanced Mach-Zehnder interferometer (MZI) is used as a frequency discriminator to calibrate the laser wavelength, which we sweep from 1550.75-1554.6~nm. Figure~\ref{Figure2}A presents an example spectrum of a ring resonator with 1300~nm waveguide width and 60~$\mu$m radius.
We identify the micro-ring cavity modes as resonance dips in the measured spectrum and separate the singlet and doublet resonances by calculating their asymmetric factors, which indicate how far the resonance peaks deviate from an ideal Lorentzian lineshape. 
Singlet resonances can be fitted using the standard Lorentzian lineshape to estimate intrinsic quality factors. Doublet resonances, however, are the result of resonance splitting in the resonator and need to be fitted using a coupled resonance theory \cite{Goro2000}.
Three fundamental TE mode resonance dips in the example spectrum are presented in Fig.~\ref{Figure2}B, labeled as mode a (singlet), mode b (doublet) and c (doublet) with extracted intrinsic Q-factors of $3.3 \times 10^5$, $4.7 \times 10^5$, and $4.1 \times 10^5$, respectively.

Figure~\ref{Figure2}C plots the mean values with their standard deviation of the intrinsic Q-factors as a function of different ring waveguide widths. 
From the measured Q-factors, we calculate the propagation losses for each ring waveguide width, depicted in Fig.~\ref{Figure2}D, using 
\begin{equation}
\mathrm{Loss} = \frac{2\pi \cdot 10\log_{10}(e) \cdot n_{\mathrm{g}}}
{\lambda \cdot Q_{\mathrm{i}}}
\; \mathrm{dB/cm},
\label{eq:loss}
\end{equation}
where $n$\textsubscript{g} is the group index for the fundamental TE mode, $\lambda$ is the wavelength, and $Q$\textsubscript{i} is the intrinsic Q-factor.
We observe an increase in the Q-factors with increasing ring waveguide widths due to the decreased modal overlap of the fundamental TE mode with the waveguide sidewalls and reduced scattering losses in the wider waveguides. 
Since the losses are inversely proportional to the intrinsic Q-factors, they show a decreasing trend for larger waveguide widths. The losses for narrow waveguides of 500~nm and 600~nm widths reach up to 8.69~dB/cm and 7.76~dB/cm, respectively, whereas the losses for waveguide widths beyond 1200~nm are well below 2~dB/cm, with the lowest propagation loss at 1.24~dB/cm.
More details on the experimental setup for ring resonator characterization and resonance analysis are provided in the Supplementary Materials.\\

\clearpage


\subsection*{Simulation and measurement of backward SBS}

\begin{figure}[ht] 
	\centering
	\includegraphics[width=0.95\textwidth]{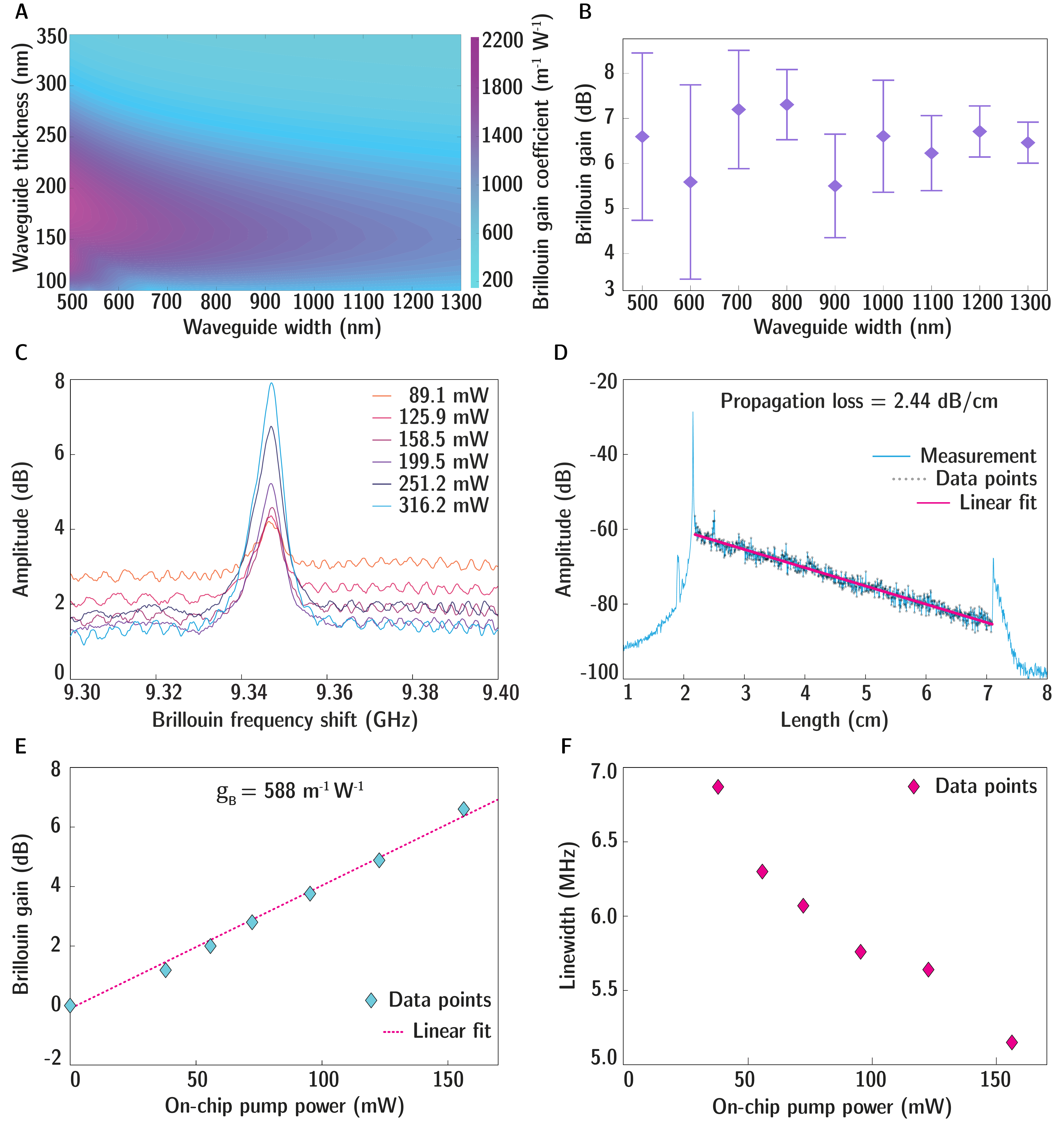} 

	\caption{\textbf{Simulation of Brillouin gain coefficients and backward SBS measured in a 49.8~mm paperclip waveguide.} (\textbf{A}) Numerical simulation of maximum Brillouin gain coefficient for different waveguide widths and heights.
    (\textbf{B}) Calculated total Brillouin gain as a function of various waveguide widths. Gain values are calculated for a waveguide height of 250~nm using the simulated gain coefficients in (A), measured ring waveguide losses, and assuming a waveguide physical length of 49.8~mm with on-chip pump power of 156~mW.
    (\textbf{C}) Brillouin peaks measured at 89.1-316.2~mW input pump powers. (\textbf{D}) Linear fitting for the propagation loss measurement. (\textbf{E}) Linear fitting for measured Brillouin gain against on-chip pump power. (\textbf{F}) Measured backward SBS linewidth at various on-chip pump powers.}
	\label{Figure3} 
\end{figure}

We use the open-source finite element software NumBAT \cite{NumBAT,8731642} to calculate the SBS response of air-cladded InGaP-on-SiO\textsubscript{2} waveguides, pumped at 1550~nm. To facilitate backward intra-modal SBS, we assume quasi-TE polarized fundamental modes for both counter-propagating pump and probe waves, a mechanical quality factor of 1800 for the calculation, and sweep the waveguide's height and width from 100-350~nm (10~nm step) and 500-1300~nm (20~nm step), respectively. 
From each calculated Brillouin spectrum, we extract the maximum Brillouin gain coefficient. Figure~\ref{Figure3}A depicts the resulting distribution indicating an increasing Brillouin gain coefficient with decreasing waveguide width, where we find the highest Brillouin gain coefficient for a waveguide with estimated height of around 180~nm and width of around 500~nm. 
The relevant material parameters for simulation can be found in the Supplementary Materials. 

Figure~\ref{Figure3}B presents the estimated Brillouin gain as a function of different waveguide widths for the waveguide height of 250 nm. The gain is calculated using
\begin{equation}
G = \exp\left( g_{\mathrm{B}} L_{\mathrm{eff}} P_{\mathrm{p}} \right),
\label{eq:gain}
\end{equation}
where $P\textsubscript{p}$ is the coupled pump power and $L\textsubscript{eff}$ is the effective waveguide length \cite{Morrison:17}, calculated using $L_{\mathrm{eff}} = \frac{1-\exp(-\alpha L)}{\alpha}$, where $\alpha$ is the propagation loss at the pump wavelength (in units of $\mathrm{cm}^{-1}$) and $L$ is the physical length of the waveguide.
For the Brillouin gain calculations we use the propagation loss measurement results from Fig.~\ref{Figure2}D, the Brillouin gain coefficient simulation results from Fig.~\ref{Figure3}A, and assume a 49.8~mm physical waveguide length and 156~mW coupled pump power.
We observe in Fig.~\ref{Figure3}B that the Brillouin gain values stay within 5-8~dB and are not significantly dependent on the various waveguide widths, due to the trade-off between the Brillouin gain coefficient and propagation losses in different waveguide geometries. Although the Brillouin gain coefficient is significantly higher for narrow waveguides, the total gain value is hampered by the significantly higher optical losses. 
On the other hand, the use of wider waveguides significantly reduces the propagation losses, hence increasing its effective length and the resulting Brillouin gain.
Therefore, we experimentally investigate backward SBS in a 1000~nm wide, 250~nm thick, and 49.8~mm long waveguide with a pump-probe setup, detailed in the Supplementary Materials.

We perform several experiments with increasing pump powers to investigate the Brillouin gain and nonlinear behavior of the SBS resonance in InGaP waveguides.
The resulting Brillouin gain spectra are depicted in Fig.~\ref{Figure3}C. We extract a Brillouin frequency shift of 9.346~GHz, which is in excellent agreement with the numerical simulation (Fig.~\ref{Figure0}D), with a slight difference of 7~MHz between calculation and experiment.
We measure the propagation loss of the SBS waveguide using an Optical Vector Analyzer (OVA) and obtain 2.44~dB/cm based on a linear fit (see Fig.~\ref{Figure3}D), resulting in an effective length of 16.7~mm for the 49.8~mm physical length.
In addition, we extract the Brillouin gain values from Fig.~\ref{Figure3}C and plot them over linear pump power in Fig.~\ref{Figure3}E, where 38-156~mW coupled powers correspond to 89.1-316.2~mW input powers. As expected for backward SBS, a linear increase of Brillouin gain (in logarithmic scale) over on-chip pump power is observed, with maximum Brillouin gain of 6.6~dB at a coupled pump power of 156~mW, matching the calculated mean value for 1000~nm waveguide width (Fig.~\ref{Figure3}B). From the linear fit and calculated effective length, we extract the Brillouin gain coefficient of 588~W\textsuperscript{-1}m\textsuperscript{-1}.
Furthermore, we estimate the Brillouin peak linewidth by fitting the resonances to a Lorentzian function. The linewidth over different pump powers are presented in Fig.~\ref{Figure3}F, showing a narrowing linewidth with increasing pump power, a typical characteristic of stimulated Brillouin scattering \cite{Morrison:17}. 
We obtain an ultra-narrow linewidth of 5.2~MHz at 156~mW, the narrowest backward SBS linewidth demonstrated in a photonic chip to the best of our knowledge.
This could be explained by the aluminum-free waveguide core material which enables ultra-low defects to ensure low phonon scattering.

\clearpage


\section*{Discussion}

In this work, we showcase the capability of InGaP-on-SiO\textsubscript{2} platform for backward SBS by measuring an ultra-narrow SBS linewidth of 5.2~MHz at a Brillouin frequency shift of 9.346~GHz. Leveraging the outlined wafer-scale fabrication, the direct bonding process opens up a wide range of heterogeneous integrated photonic circuits based on III-V-on-insulator, providing high optical nonlinearities and strong confinement. 

\begin{figure}[ht] 
	\centering
	\includegraphics[width=0.95\textwidth]{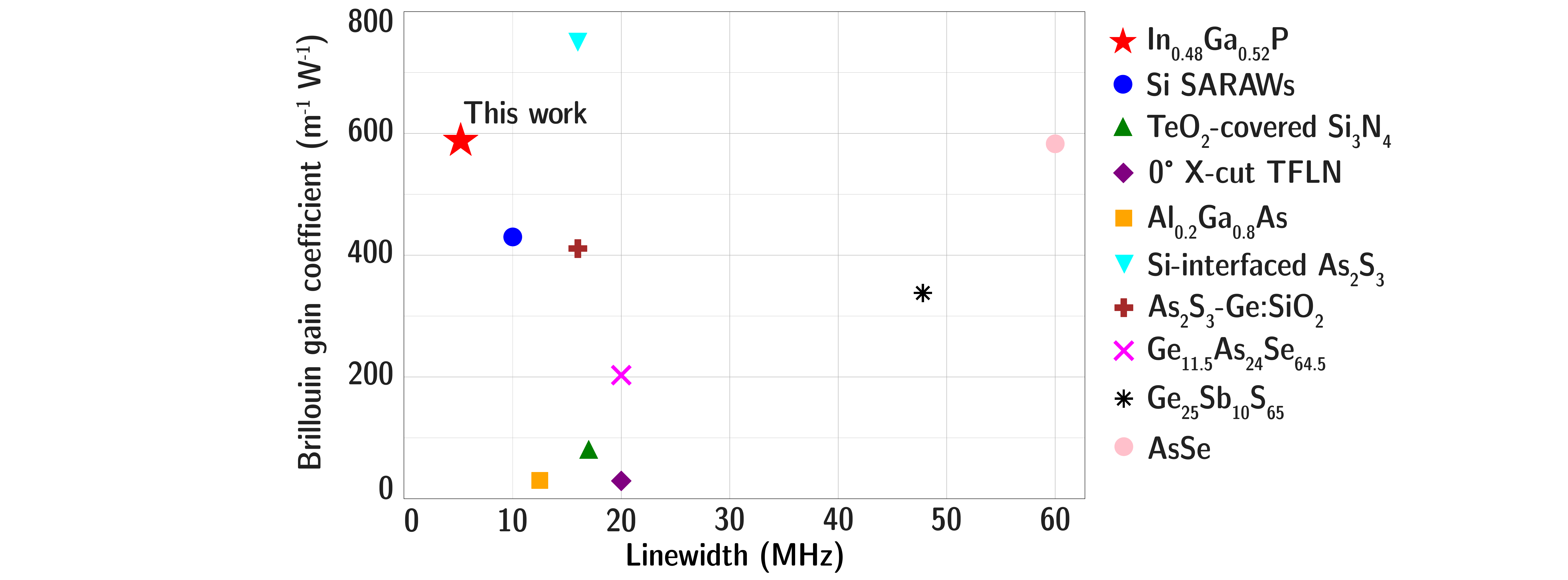} 
	\caption{\textbf{Backward SBS gain coefficient against narrowest linewidth measured for various material platforms.} Comparison of experimentally measured backward SBS gain coefficient and corresponding narrowest linewidth for the InGaP on insulator material platform and previously reported waveguide materials and structures, including silicon suspended anti-resonant acoustic waveguides (SARAWs) \cite{Lei2024}, tellurite-covered silicon nitride \cite{Klaver2026}, $0^\circ$ x-cut TFLN \cite{doi:10.1126/sciadv.adv4022}, AlGaAs on insulator (horizontal
shear wave) \cite{Jin:26}, silicon interfaced As\textsubscript{2}S\textsubscript{3} \cite{Morrison:17}, hybrid chalcogenide-germanosilicate \cite{https://doi.org/10.1002/adfm.202105230}, GeAsSe glass \cite{10.1063/5.0220496}, GeSbS chalcogenide \cite{9431687} and arsenic selenide planar waveguides \cite{Lai:23}.
}
	\label{Figure4} 
\end{figure}

Figure~\ref{Figure4} presents a comparison of the state-of-the-art SBS material platforms with their respective narrowest backward SBS linewidth. We recognize that the other materials cannot achieve a linewidth below 10~MHz, whereas InGaP breaks through this barrier with a linewidth of 5.2~MHz (factor of two improvement), alongside a notable Brillouin gain coefficient of 588~W\textsuperscript{-1}m\textsuperscript{-1}, competitive against the suspended-Si waveguides, ChGs and Si-ChG hybrid waveguides, and one order of magnitude higher than other material platforms. Hence, we anticipate a phonon storage time of 61.2~ns in an InGaP-on-insulator photonic-phononic memory, matching its acoustic decay time \cite{Merklein2017}.

The current loss of 2.44~dB/cm limits the achievable gain, and we expect to reduce the loss to around 0.6~dB/cm with an InGaP passivation process \cite{Ahler:26}. We can further improve the Brillouin gain coefficient by engineering the waveguide geometry: based on current simulation results (Fig.~\ref{Figure3}A), reducing the InGaP thickness to 160~nm for 1000~nm width will increase the Brillouin gain coefficient by a factor of 2.5, making InGaP-on-SiO\textsubscript{2} a leading platform for future Brillouin-enabled photonic devices. Additionally, we can combine high-Q ring resonators with high-gain SBS waveguides to generate a Brillouin-Kerr frequency comb \cite{PhysRevLett.126.063901}, utilizing the strong $\chi$\textsuperscript{(3)} nonlinearity in InGaP.

Owing to its outstanding Brillouin spectral characteristics, InGaP-on-insulator paves the way for next-generation microwave photonic applications, including narrow linewidth filters, lasers, acousto-optic memories, sensors and isolators \cite{Eggleton2019-fa,Lai:25}. Since III–V semiconductors are widely used as active components in integrated photonics as they provide direct bandgaps and strong electro-optic effects, we could potentially embed integrated lasers, quantum sources, optical modulators and acoustic wave amplifiers into the InGaP-on-insulator platform, enabling system-level operations such as acousto-electric amplification of Brillouin gain \cite{1395890,Senellart2017,DAVENPORT2018139,Hackett2021,PhysRevApplied.19.014059}.

\clearpage


\section*{Materials and Methods}

\textbf{Fabrication details}\\
The wafer-scale fabrication process starts with molecular beam epitaxy growth of the 250~nm InGaP device layer on a 200~nm thick Al\textsubscript{0.8}Ga\textsubscript{0.2}As etch-stop layer on a GaAs handle wafer. After the epitaxial growth, the fabrication process continues  with patterning cross-shaped fill patterns in the thermal oxide layer of an oxidized silicon wafer (Fig.~\ref{Figure1}A, step 1). The cross-shapes have overall height and width of 50~$\mu$m, and arm width of 3~$\mu$m. Such structures are required to trap gases generated during the subsequent wafer-scale bonding of the epitaxial InGaP layer on the oxidized silicon wafer. Prior to bonding, we deposit 10~nm and 2~nm alumina interlayers on the InGaP and oxidized-Si wafers, respectively, using atomic layer deposition (ALD). Next, we activate both wafers with O\textsubscript{2} plasma, followed by direct bonding and a $200~^\circ$C annealing step to form a permanent bond. After the bonding process, we chemically remove the GaAs substrate and the etch-stop layer in NH\textsubscript{4}OH:H\textsubscript{2}O\textsubscript{2} solution and dilute HF, respectively \cite{Chiles:19} (Fig.~\ref{Figure1}A, step 2). To form the optical waveguides, we deposit 150~nm SiO\textsubscript{2} hard mask through plasma enhanced chemical vapor deposition (PECVD), followed by hard mask patterning with electron-beam lithography (EBL) and dry etch using CHF\textsubscript{3} plasma chemistry (Fig.~\ref{Figure1}A, step 3).
Subsequently, we dry-etch the InGaP waveguide structures using BCl\textsubscript{3} chemistry and remove the SiO\textsubscript{2} hard mask in dilute HF (Fig.~\ref{Figure1}A, step 4). Finally, we release the 1~$\mathrm{cm}^2$ square chips through facet-etching and deep-silicon-etching processes.
\\
\\
\textbf{Chip design details}\\
The first chip design consists of straight waveguides with a fixed length of 9.8~mm while ranging the widths from 300-1000~nm in 10~nm step size, used for fab-quality control. We cleave one of the chips in the middle and scan the cross-section for all straight waveguides. 
In the second chip design, we introduce paperclip waveguides with 20~$\mu$m bend section radius to facilitate various waveguide lengths ranging from 9.8-99.8~mm, for widths in between 300-1000~nm (100~nm step). Figure~\ref{Figure1}D shows paperclip structures with meandering waveguides. Such design enables realization of waveguides of different lengths in a compact, 1~cm wide, chip with the 20~$\mu$m bend radius designed to minimize bend losses.
The third chip design includes waveguide-coupled ring resonators of various widths. Here, we measure the resonator optical quality factors as a function of different ring waveguide width to estimate the optical propagation loss in the InGaP-on-SiO\textsubscript{2} platform. The straight bus-waveguides have a fixed length of 9.8~mm and a 500~nm width, while the radius and width of ring waveguides are between 20-60~~$\mu$m (20~$\mu$m step) and 500-1300~nm (100~nm step size), respectively. The gap between straight and ring waveguides ranges from 100-300~nm (50~nm step). Figure~\ref{Figure1}E shows a ring resonator with 40~$\mu$m radius and a ring-bus gap of 300~nm, as well as the zoomed-in view of the coupling region.

\clearpage






\clearpage 

%
\bibliography{InGaP_SBS_converted} 
\bibliographystyle{sciencemag_allauthors}

%
%
%
%
%
%


\section*{Acknowledgments}

We thank Peter Lowell and Alvin Flores of NIST Boulder micro-fabrication facility for useful discussions and cleanroom tool support during the process development and device fabrication. We also thank Zixuan Wang, Joey Bush, Karthik Myilswamy and Nathan Wriedt for useful discussions.

\paragraph*{Disclaimer:} This document has not been peer reviewed but has been cleared by NIST for release. Any mention of commercial products is for information only; it does not imply recommendation or endorsement by NIST. 

\paragraph*{Funding:}

This work has been supported through the NIST-on-a-Chip (NoaC) program and the Australian Research Council (Grant Nos. DP220100488, DE230100964, and CE230100006).
NIST work was funded solely by the United States Government.

\paragraph*{Author contributions:}
Conceptualization: Y.X, A.B, and N.N. 
Fabrication and growth: Y.X, K.L.S and N.N. 
Experimental investigation: Y.X, L.H, R.L.R, C.K.L, M.M and N.N. 
Numerical simulation: Y.X, L.H, M.J.S and N.N. 
Visualization: Y.X, L.H. 
Supervision: L.H, G.S, A.B and N.N. 
Project administration and funding acquisition: A.B and N.N. 
All authors contributed to the writing of the manuscript and discussion of the results.

\paragraph*{Competing interests:}

The authors declare that they have no competing interests.

\paragraph*{Data and materials availability:}
All data needed to evaluate and reproduce the results in the paper are present in the paper, the Supplementary Materials, and a publicly accessible online repository.

\clearpage

\subsection*{Supplementary Materials}

SBS experimental details\\
Intrinsic Q-factor analysis for ring resonators\\
Material parameters for NumBAT simulation\\
Figs. S1 to S3\\
Table S1\\
References \cite{Suzuki_1983, Schneider:19}\\


\newpage



\end{document}